\documentclass{aa}
\usepackage{psfig}
\begin{document}

\title{On the Origin of the FeK$\alpha$ Line in the Seyfert 2 Galaxy NGC 7172}

   \author{M. Dadina
              \inst{1,2,3}
	\and  L. Bassani
              \inst{3}
	\and M. Cappi
	      \inst{3}
	\and  G.G.C. Palumbo
	      \inst{4}
	\and L. Piro
	      \inst{2}
	\and M. Guainazzi
	      \inst{5}
        \and G. Malaguti
	      \inst{3}
	\and G. Di Cocco
	      \inst{3}
	\and M. Trifoglio
              \inst{3}
}

   \offprints{M. Dadina (dadina@tesre.bo.cnr.it)}

%1
    \institute{BeppoSAX SOC, Telespazio, via Corcolle 19, I-00131, 
               Roma, Italy
%2
	      \and Istituto di Astrofisica Spaziale, CNR, via Fosso del Cavaliere, I-00133, 
               Roma, Italy
%3
	      \and Istituto Te.S.R.E, CNR, via Gobetti 101, I-40129, Bologna, Italy
%4
	      \and Universit\`a degli Studi di Bologna, Dip. di Astronomia, Via Ranzani 1, I-40127, Bologna, Italy 
%5
	      \and XMM-Newton SOC, VILSPA - ESA, Apartado 50727, E-28080, Madrid, Spain
}

 \date{Received / Accepted}

\titlerunning{The origin of FeK$\alpha$ line in NGC 7172}
\authorrunning{M. Dadina et al.}

 \abstract{
The Seyfert 2 Galaxy NGC 7172 was observed twice by BeppoSAX narrow field 
instruments approximately one year apart. 
As found in previous observations, the source is variable on short 
time scales (hours) by a factor of $\sim$30$\%$, confirming the presence of a 
type 1 nucleus at the center. 
A strong flux variability (by a factor 
of 2) on a longer time scale 
was observed between the two observations. Indeed, in November 1997, 
the lowest flux ever recorded was detected by BeppoSAX. 
The broad band spectra obtained with the BeppoSAX narrow field 
instruments show  marginal evidence of a reflection component. 
This structure could explain the very flat spectra previously observed
in the 2-10 keV band.
An FeK$\alpha$ line was also detected in both observations. 
The line intensity appears to remain almost constant between the two 
observations even if the 
associated large errors cannot exclude some variability. In order to understand
the problem of the origin of the line, data from previous observations, 
performed with other satellites, were also used. 
The scenario favored is one where the line is produced at a distance
of approximatively  8 light years from the continuum source region, 
rather than having its origin in the accretion disk.  In the 
framework of the standard unified models for AGN, this suggests  
that its origin is located in the molecular torus.
}

\maketitle

\section{Introduction}

NGC 7172, one of the prototype narrow emission line galaxies (NELGs, Bradt et al. 1978), is a
 S0-Sa type galaxy  (z=0.00868)  member  of the
Hickson's compact group HCG90. The galaxy  is seen edge-on and has a
prominent  equatorial dust lane. Its nuclear optical emission
spectrum is
indicative of excitation by a power-law photoionization source.
The large narrow Balmer line decrement ($\geq$6.5, Sharples et al. 1984)
indicates substantial obscuration in the narrow line region (A$_{v}$$\sim$
2.5 mag or 5.6$\times$10$^{21}$ cm$^{-2}$).

The galaxy, first identified as
an X-ray source by the HEAO1/A2 experiment (Marshall et al. 1979)
is one of the brightest Seyfert 2 galaxies at these energies and one
of the Piccinotti sample sources (Piccinotti et al. 1982). It has been 
the target of various observations in both soft and hard X-rays since
its discovery
(see Polletta et al. 1996 for a recent compendium of X-ray data).

At soft energies, the location of the source in the compact group
complicates the spectral analysis and recent ROSAT data suggest
that most of the emission below 2 keV is indeed due to the intergalactic
gas of the group (Guainazzi et al. 1998).

The source is variable in the 2-10 keV range both on short and long
time scales (Guainazzi et al. 1998); in particular, after
a long period (1977-1995) of fairly constant flux, the source suddenly dimmed 
by almost
a factor of 4 within 5 months. The spectrum in this band is fairly
simple, being characterized by a strongly 
(N$_{H}$$\sim$8$\times$10$^{22}$ cm$^{-2}$) 
absorbed power-law  (photon index $\Gamma$=1.5-1.8) and a narrow line at 6.4 
keV  with an equivalent width of 40-120 eV (Smith and Done 1996, 
Turner et al. 1997, Guainazzi et al. 1998).

During the latest ASCA observation (Guainazzi et al. 1998),
characterized by a low flux level ($\sim$10$^{-11}$erg cm$^{-2}$ s$^{-1}$), 
the spectrum
was flatter and the line equivalent width (EW) higher than in previous 
observations.
The EW increase was consistent either  with a change in the
continuum and line level  or in  the continuum flux only.
Although the Ginga data gave some evidence for the presence of
a reflection component (Nandra \& Pound 1994, Smith \& Done 1996),
which could also explain the flat
spectrum observed at low flux levels, this issue is still
debatable. Alternatively, the flat spectrum could be due
to complex absorption in the source (Guainazzi et al. 1998).
NGC 7172 was also observed at high energies both by
OSSE and BATSE on CGRO: the spectral photon index at these energies
is canonical ($\Gamma$=2.0$^{+0.4}_{-0.3}$) and the flux is typically
7-8$\times$10$^{-11}$ erg cm$^{-2}$ s$^{-1}$ in the 20-100 keV band
(Ryde et al. 1997, Malizia et al. 1999).

Here we present   the broad band (0.1-100 keV) spectrum of NGC 7172
obtained by BeppoSAX on two subsequent observations
performed approximately one year apart.

\section{Observations and data reduction}

\begin{table}[h]
\caption{BeppoSAX observations log.}
\tabcolsep=3.0mm
\begin{center}
\scriptsize
\begin{tabular}{l c c }
\hline\hline
&&\\
Obs. (date)&Integration Time (ks.) & 10$^{-3}$ Counts s$^{-1}$\\
\hline
&&\\
A (Oct. 15, 1996)&&\\
&&\\
LECS&15.2 &8.9$\pm0.8$\\
&&\\
MECS&39.2&144$\pm2$\\
&&\\
PDS&17.3 &338$\pm36$\\
&&\\
\hline
&&\\
B (Nov. 6, 1997)&&\\
&&\\
LECS&22.5&6.1$\pm0.5$\\
&&\\
MECS&49.4& 51$\pm1$\\
&&\\
PDS&21.1&199$\pm33$\\
&&\\
\hline
\hline
\end{tabular}
\end{center}
\end{table}

The BeppoSAX X-ray observatory (Boella et al. 1997) is a major
program of the Italian Space Agency with participation of
the  Netherlands Agency for Aerospace Programs and ESA.
The work presented here concerns observations performed
with the Narrow Field Instruments (NFI)
on-board the satellite: the Low Energy Concentrator Spectrometers
(LECS, covering the 0.1-10 keV band; Parmar et al. 1998), the Medium Energy 
Concentrator Spectrometer (MECS, covering the 1-10.5 keV band; Boella et 
al. 1997), the High Pressure Gas Scintillation Proportional Counter (HPGSPC
covering the 4-120 keV band; Manzo et al. 1997) and the Phoswich Detection 
System (PDS, covering the 15-200 keV band; Frontera et al. 1997)

BeppoSAX NFI pointed at NGC 7172 on October 15th, 1996 (observation A)
and re-observed the source on November 6th, 1997 (observation B) because
the first pointing was
interrupted by a gamma-ray burst TOO. The second observation
was performed with only  MECS units 2 and 3 
since the unit 1 failed between the two pointings.

\begin{figure}
\begin{center}
\psfig{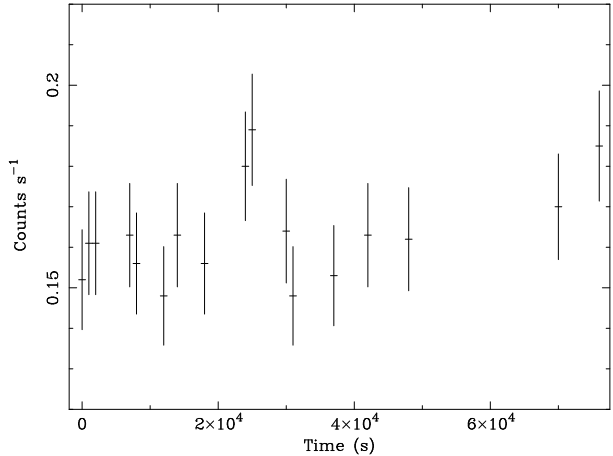}
\caption{MECS (2-10 keV) light curve of observation A (using a time bin of 1000 s).}
\end{center}
\end{figure}

\begin{figure}
\psfig{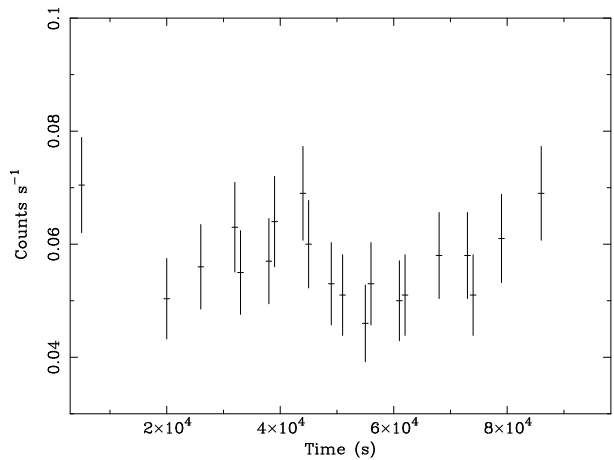}
\caption{MECS (2-10 keV) light curve of observation B (using a time bin of 1000 s).}
\end{figure}

The effective exposure times and net count rates for each
narrow field instrument are listed in Table 1 for both
BeppoSAX observations. 

For the data reduction and analysis, performed using ftools 4.1
and XANADU 10.0 software packages, we used standard
techniques and criteria.
Light curves and spectra were extracted from a region
centered on NGC 7172 and having radius of 8' for LECS and 4' for MECS.

Figures 1 and 2 show the light curves in the 2-10 keV band  for both
observations. While during observation A the source
intensity was stable except for a rapid rise and decline of 20-25$\%$
at {\it t}$\sim$2.5$\times$10$^{4}$ s, during observation B NGC 7172 
showed
a systematic variability of $\sim$30$\%$  similar
to that detected in the last ASCA observation (Guainazzi et al. 1998). This lead us to conclude that a Seyfert 1-like nucleus is present in NGC 7172. Due to the limited statistics available,  we find no significant  
spectral variations and therefore the source data were integrated over the 
entire observation to obtain average spectra.

The LECS field is certainly contaminated by  the soft X-ray emission
from the embedding compact group, while the hard X-ray band (MECS) is  
dominated by NGC 7172. No bright sources emitting in the 2-10 keV band 
are known to be present in the PDS field
of view of 1.3  degrees (FWHM); the BL Lac object PKS2155-304
is 1.8 degrees away from NGC 7172, far enough to exclude contamination
of the high energy emission.

The LECS background spectrum was extracted from
two semi annuli in the instrument field of view which were
then normalized
to the source extraction region as described in Parmar et al. (1999).
The MECS background subtraction was instead
performed by means of blank sky deep field exposures
accumulated by the BeppoSAX Science Data Center at the beginning of the
mission.
The PDS reduction was performed using the SAXDAS software package
taking into account
both rise-time and spike corrections.
Products were obtained by plain subtraction of the
``off'' from the ``on'' source data. No HPGSPC data will
be considered in the present paper as the source is too faint
for a correct background subtraction.

\begin{figure}[h]
\psfig{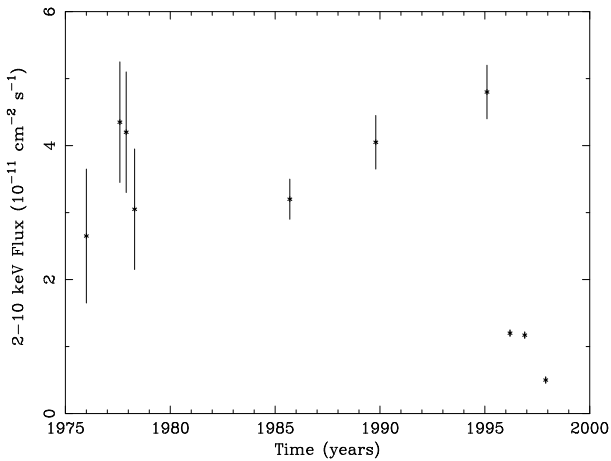}
\caption{The long-term NGC 7172 light curve. Data are from the Ariel V observation (1976), the HEAO-1/A2 observations (1977.6, 1977.9, 1978.3), the EXOSAT observation (1985.7), the Ginga observation (1989.8), the ASCA observations (1995.5, 1996.5)  and the BeppoSAX observations (1996.7, 1997.8). The source flux declines by a factor $\sim$10 from the 1995 ASCA observation to the 1997 BeppoSAX observation, during which the source was detected at the lowest state ever recorded.}
\end{figure}

LECS and MECS data were rebinned in order to sample the energy resolution
of the detector with an accuracy proportional to the count rate; furthermore,
we ensure 
at least 20 counts per channel which allows applicability of the 
$\chi^{2}$ statistics.

\begin{table*}
\title{power-law models}
\vglue0.3truecm
{\hfill\begin{tabular}{l l c c c c c c c }
\hline\hline
&&&&&&\\
Obs.$^{a}$&Mod. & N$_{H}$$^{b}$&$\Gamma$$^{c}$&E$_{Line}$$^{d}$&I$_{Line}$$^{e}$&$\chi$$^{2}$/d.o.f\\
&&&&&&\\
   & & 10$^{22}$cm$^{-2}$ & & keV &10$^{-5}$ ph. cm$^{-2}$ s$^{-1}$ &  \\
&&&&&&\\
\hline
A&abs. power-law & 10.21$^{+0.80}_{-0.79}$& 1.70$^{+0.12}_{-0.12}$& & &86.16/84\\
&&&&&&&&\\
&abs. power-law + Gaussian line& 9.84$^{+0.76}_{-0.71}$& 1.67$^{+0.09}_{-0.08}$&6.46$^{+0.16}_{-0.15}$   &   2.39$^{+1.28}_{-1.29}$ & 77.14/82\\
&&&&&&\\
\hline
\hline
&&&&&&\\
B&abs. power-law    &9.53$^{+1.38}_{-1.27}$ &1.62$^{+0.24}_{-0.14}$& & &72.33/64\\
&&&&&&\\
&abs power-law + Gaussian line&9.01$^{+1.38}_{-1.28}$&1.60$^{+0.14}_{-0.13}$&6.46$^{+0.15}_{-0.13}$&2.18$^{+1.08}_{-1.08}$&61.47/62\\
&&&&&&\\
\hline
\hline
\end{tabular}\hfill}
\caption{ ($^{a}$)BeppoSAX observation, ($^{b}$)absorbing column density, ($^{c}$)photon index, ($^{d}$)line energy, ($^{e}$)line intensity.}
\vspace{0.3cm}
\end{table*}

\begin{table*}
\title{power-law + reflection component + Gaussian line}
\vglue0.3truecm
{\hfill\begin{tabular}{l l c c c c c c}
\hline\hline
&&&&&&&\\
Obs.$^{a}$ &N$_{H}$$^{b}$&$\Gamma$$^{c}$&E$_{Line}$$^{d}$&I$_{line}$$^{e}$& R$^{f}$ &$\chi$$^{2}$/d.o.f\\
&&&&&&\\
&10$^{22}$cm$^{-2}$&&keV&10$^{-5}$ ph. cm$^{-2}$ s$^{-1}$&&\\
&&&&&&\\
\hline
&&&&&&\\
A&10.18$^{+0.80}_{-0.78}$&1.91$^{+0.23}_{-0.21}$&6.45$^{+0.21}_{-0.33}$&1.72$^{+1.40}_{-1.50}$&2.24$^{+3.96}_{-1.61}$&73.64/81\\
&&&&&&\\
\hline
\hline
&&&&&&&\\
B &9.82$^{+1.38}_{-1.41}$&1.91$^{+0.41}_{-0.34}$&6.47$^{+0.20}_{-0.15}$&1.92$^{+1.16}_{-1.21}$&2.11$^{+0.97}_{-1.95}$&58.61/61\\
&&&&&&\\
\hline
\hline\end{tabular}\hfill}
\caption{ ($^{a}$)BeppoSAX observation, ($^{b}$)absorbing column density, ($^{c}$)photon index, ($^{d}$)line energy, ($^{e}$)line intensity, ($^{f}$)relative reflected-direct normalization.}
\vspace{0.3cm}
\vspace{0.3cm}
\end{table*}

\section{Spectral Analysis}

\begin{figure}[h]
\psfig{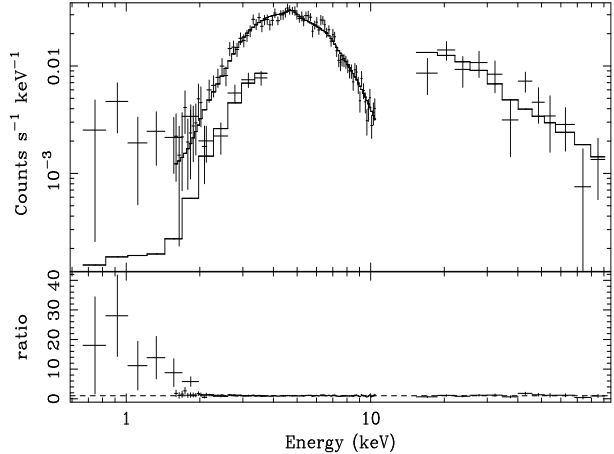}
\caption{The 0.6-90 keV BeppoSAX data of observation A modeled with a simple absorbed power-law (upper panel). In the lower panel the ratio between the data and model is plotted. A huge soft excess is clearly visible at low energies extending from 0.6 up to $\sim$ 2.0 keV. Very similar results are obtained for observation B.}
\end{figure}

\begin{figure}[h]
\psfig{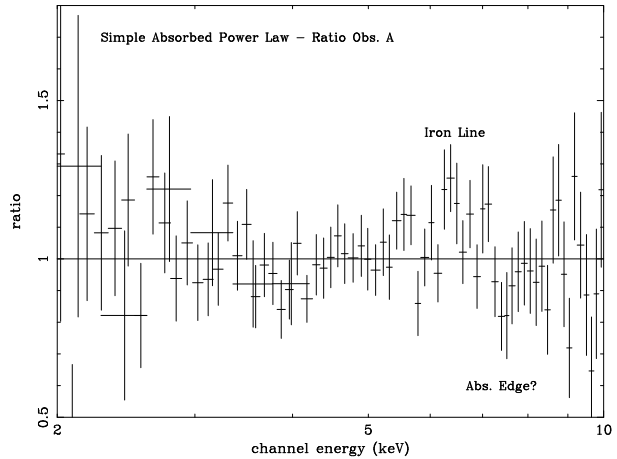}
\caption{The ratio between the data and the simple power-law model in the 
2-10 keV band for observation A. The features due to the FeK$\alpha$ fluorescent line (at 6.4 keV) and, possibly, an absorption edge (at 7.1 keV) are visible.}
\end{figure}

\begin{figure}[h]
\psfig{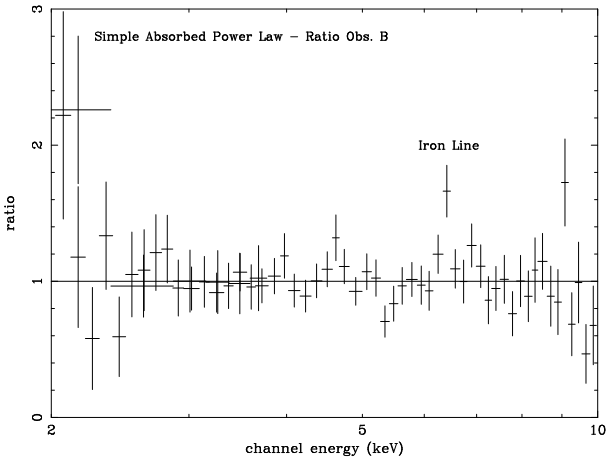}
\caption{The ratio between the data and the simple power-law model in the 
2-10 keV band for observation B. A soft excess extending in the 2-2.5 keV band is visible.}
\end{figure}

Spectral data from LECS, MECS and PDS were fitted simultaneously.
We restricted the spectral analysis to the 0.12-4.0 keV range for LECS
and 1.65-10 keV energy bands for MECS
where the response matrices released in September 1997
are best calibrated; PDS analysis was restricted to data below
90 keV as the source signal disappears above this energy.
Normalization constants were introduced to allow for known
differences in the absolute cross-calibration between detectors.
The values of the two constants were allowed to vary. The MECS-LECS
constant always turned out to be within 10$\%$ of the suggested value of 0.9
(see Fiore, Guainazzi \& Grandi 1999 for a discussion). 
The MECS-PDS constant was  constrained to vary within the suggested 
0.7-0.9 interval (Fiore, Guainazzi \& Grandi 1999).

All quoted errors
correspond to 90$\%$ confidence intervals for one interesting parameter
($\Delta\chi$$^{2}$=2.706).
All models used in what follows contain an additional term to allow
for the absorption of X-rays due to our Galaxy which, in the direction of
NGC 7172, is 1.65$\times$10$^{20}$ cm$^{-2}$ (Dickey \& Lockman 1990).

During observation A, the  2-10 keV flux was 
$\sim$1.2$\times$10$^{-11}$ erg cm$^{-2}$ s$^{-1}$, i.e similar to what was 
observed
in the ASCA observation performed only 5 months earlier.
In observation B, the flux decreased to $\sim$5$\times$10$^{-12}$ erg cm$^{-2}$
s$^{-1}$, indicating that the long trend of intensity decline (see figure 3)
 firstly 
observed by ASCA in the 1996 observation, was still persisting a year later.

The broad band spectra were first fitted with a simple
absorbed power-law (figure 4) to highlight any extra features.
The most prominent deviation from an
absorbed power-law is a huge soft excess below $\sim$2.0 keV.
Due to the low statistic available in
this energy range, no detailed investigation about this band of the spectrum 
is possible.
Nevertheless, following Ponman et al. (1996) and Guainazzi et al. (1998), we 
tried to fit the excess 
with a thermal plasma model (namely the MEKAL model in xspec, Mewe, 
Gronenschild \& van der Oord 1985, Mewe \& van der
 Oord 1986, Kaastra 1992, Arnaud \& Rothenflug
1985), adopting, for 
energies $\geq$2 keV, the best fit model discussed in the 
next section. The 
absorbing column density associated with this component was fixed to 
the galactic value while the metal abundance was assumed to be 
0.2 as obtained by ROSAT (Ponman et al. 1996). This model provides a best fit 
temperature of 0.9$\pm$0.2 keV and 1.2$\pm$0.5 keV for observation A and B 
respectively. These values are, within the errors, in agreement with the 
temperature measured by ASCA (0.68$\pm$0.12 keV, Guainazzi et al. 1998). 
As already suggested, this soft component is probably due to the thermal 
emission of the gas in the compact group.

\subsection{Emission in the 2-90 keV energy range}

To avoid contamination from the compact group diffuse emission, we restricted
the spectral analysis of the nuclear emision of NGC 7172 to data above 2 keV for observation 
A and 
2.5 keV for observation B. We expect that the effects of the thermal plasma associated with 
HCG90 are negligible at these energies. 

Fitting the data in these restricted energy bands with an absorbed power
law model provides a column density of $\sim$10$^{23}$ cm$^{-2}$ and a 
photon
index of 1.6-1.7 (table 2). The photon index is in 
agreement with the values obtained for samples of Seyfert galaxies 
observed both by Exosat and Ginga ($\Gamma$$\sim$1.7-1.8, Turner \& Pounds 
1989; Nandra \& Pounds, 1994) but it is only marginally consistent with what 
was
obtained in the ASCA 1995 and 1996 observations, when the source showed a very 
flat spectrum ($\Gamma$$\sim$1.5, Ryde et al. 1997; Guainazzi et al. 1998).

Inspection of figures 5 and 6, where the ratio between the data and the
absorbed power-law model is plotted, indicates the presence of
an FeK$\alpha$ line and possibly (in observation A) an absorption edge
in the 6-8 keV band.

\begin{figure}[h]
\psfig{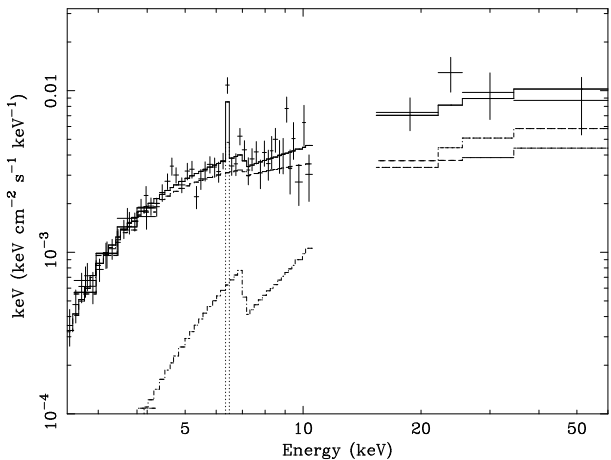}
\caption{BeppoSAX spectrum of NGC 7172 (observation B) fitted with a power-law plus a reflection component and a narrow Gaussian line (PDS data have been rebinned so as to have S/R $\geq$3 for clarity).}
\end{figure}

\begin{figure}[h]
\psfig{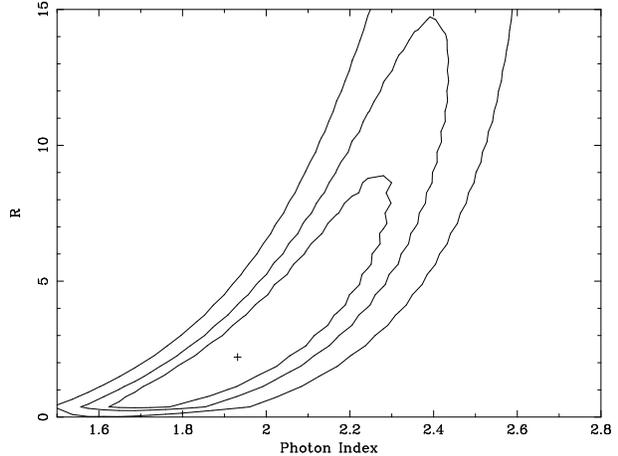}
\caption{The relative reflected-direct normalization ($R$) vs. the photon index $\Gamma$ contour plot for observation B. The contours are at 68, 90 and 99\% confidence levels going from the inner to the outer. The model is acceptable with a confidence level of 90\%. A completely similar plot is obtained for observation A.}
\end{figure}

The addition of a narrow line to the absorbed power-law model
provides an improvement in the quality of the fit  by a
$\Delta$$\chi$$^{2}$$\sim$9.0
for two extra degrees of freedom for both 
observations A and B, implying that the line
is required at more than 99.9$\%$ confidence level (table 2).
The line is located at 6.46$\pm$0.20 keV and has an equivalent width (EW)
of 120$\pm$65 eV in the first observation, increasing to 
210$\pm$105 eV in the second one. If the line width is allowed to vary,
the additional parameter does not give a significant 
improvement in the quality of the fit and yields an upper limit to the
Gaussian width of $\sigma$$\leq$0.75 keV, consistent with the energy 
resolution of the MECS (7$\%$ FWHM at $\sim$6.4 keV, Piro et al. 1995, Boella et al 1997). Due to the available statistics,
the line width was thus frozen to zero in the subsequent analysis.

Due to the possible presence of an absorption edge in observation A (no similar
feature is present in observation B), we added this component
to the previous model. The energy of the edge was 
found to be 7.3$\pm$0.5 keV but it was not statistically significant
($\Delta$$\chi$$^{2}$=3.8 for 2 more parameters of interest) giving an 
optical depth $\tau$$\leq$0.4 ($\tau$=N$_{H}$$\times$$\sigma(E)$ where 
$\sigma(E)$ is the photoelectric cross section).

Since Guainazzi et al. (1998) used a dual absorber model to explain the flat
 spectrum for the ASCA
1996 data, we tried a similar model. The quality of the data prevents us from 
constraining the 
values of the model parameters if they are allowed to vary. For this reason 
and also for 
comparison with the ASCA 1996 results, the power-law photon index was fixed to 1.9 (as obtained by 
Guainazzi et al. 1998) and the FeK$\alpha$ line energy was  
frozen to 6.4 keV. We obtained a slight improvement in the fit and the two 
column densities 
became 10$^{23}$cm$^{-2}$ for the total absorber and 
6$\times$10$^{23}$cm$^{-2}$ for the 
partial absorber with a covering fraction of $\sim$30\%. 
However, again there was no significant improvement in the fit and we had no 
evidence
either for a flat spectrum as observed by ASCA in 1996 or for an absorption 
edge. We, 
therefore, do not consider this model necessary for our data and do not 
discuss it further.

Instead, we consider the possible presence in NGC 7172 of a reprocessed 
component due to reflection  in an optically thick cold medium 
(table 3).  This is a common feature in 
Seyfert galaxies and we used the pexrav model in xspec (Magdziarz \& Zdziarski 1995) to model it.  Due to the poor statistics, we fixed the inclination angle of the reflector to be 60 deg, a value which is expected to be suitable for the 
Seyfert 2 galaxies in the framework of the standard unified models.
When this component is added to the absorbed power-law plus emission line 
model, the fit improves by $\Delta$$\chi$$^{2}$=3.5 for observation A and 2.9
 for observation B (figure 7) for only one more extra parameter of 
interest.
An F-test indicates that this component is required with a confidence level 
higher than 90$\%$ for both observations.
 Evidence for the presence of a reflected component in the 
continuum was also found in Ginga data.
 The $R$ parameter of the fit, that indicates the relative amount of 
reflection compared to the directly-viewed primary power-law, is 
only poorly constrained (figure 8) giving a best fit value of $\sim$ 2 for both
observations.  It is not clear if the reflection component is an 
artifact of the fitting procedure. Given the low 
statistics, we cannot exclude that this strong reflection is, at least in 
part, due to an artifact of the fitting procedure (namely, the 
interplay of $R$ with the MECS-PDS cross calibration factor). Assuming that it
is indeed real and that the accretion disk is actually the reflector, we would
expect a value of $R$ constant in time (since the disk reflection would follow
the continuum variations). This is
consistent with the fit results (table 3). However, associated with this 
component, we would also expect to see a FeK$\alpha$ line with constant 
EW$\sim$300 eV (George \& Fabian 1991). As will be discussed in the next 
section, this is not the case for NGC 7172, at least for observation A. On the 
contrary, assuming that the reflection
component is produced far from the direct continuum emission region, we would
expect an increase of the parameter $R$ in association with a reduction of
the source luminosity, but the quality of the data prevents us 
from verifying this point. Also, considerations of the photon index 
$\Gamma$ for 
the simple absorbed power-law model are not of much help.  
The spectrum should harden from observation  A to 
observation  B but again this is only marginally observed (see table 2). 
A possible indication of the absence of this structure in the 
source's spectrum comes from the RXTE observations of NGC 7172 that do not 
require reflection to fit the data well. When this component is added to a 
simple absorbed power-law plus 
Gaussian line model, a marginal improvement of the fit is obtained with 
$\Delta$$\chi$$^{2}$$\sim$0.5-1.0 for one more extra parameter 
(Georgatopoulos \& Papadakis 2000).

Finally, it is worth noting that, whatever the model used, 
the absorbing column density turns out to be N$_{H}$ $\sim$10$^{23}$cm$^{-2}$, 
well in agreement with what has been observed previously. 
This implies that the 
state of the absorber is not affected by the state of the source.

\subsection{The  FeK$\alpha$ Line}

The  FeK$\alpha$ line in NGC 7172 was first detected by ASCA in the 1995 observation 
(Ryde et al. 1996, Guainazzi et al. 1998), while a previous Ginga observation 
in 1989 provided only an upper limit of $\sim$40 eV;
BeppoSAX detected the line in both observations. 
Fitting the continuum with an absorbed power-law, we obtain an 
equivalent width of 120$\pm$65 and 210$\pm$105 eV for observation A and B 
respectively, further indicating that the line EW increases as the 
source flux decreases.

\begin{figure}[h]
\psfig{file=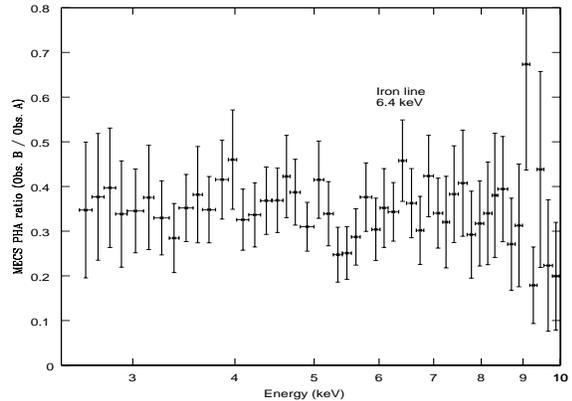,height=6cm,width=8.0cm,angle=0.0}
\caption{MECS PHA ratio between the two BeppoSAX observations (B/A). 
No significant excess is visible at the energy of the FeK$_{\alpha}$ line.}
\end{figure}

Figure 9 shows the MECS PHA ratio between the two BeppoSAX observations 
(B/A).
In the plot, no significant feature is present at the energy of the 
FeK$\alpha$ line. At first glance, this would imply that the line follows the 
changes of 
the continuum flux.  
However, a more careful inspection indicates that the absence of any 
significant feature in the PHA ratio is possibly due to our limited statistics.
Assuming that the line flux was constant between the two observations, we 
predict a $\sim$10\% feature in the PHA ratio,  {\it i.e.} an undetectable excess considering the statistics available.

\begin{figure}[h]
\hspace{0.37cm}\psfig{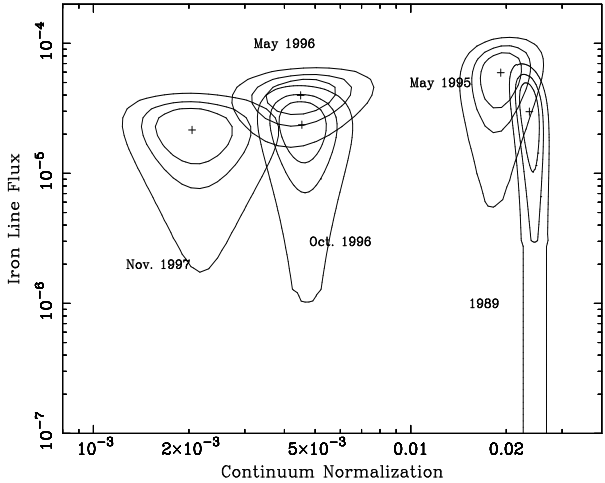}
\vspace{0.3cm}
\caption{Contours of the line flux (units of ph. cm$^{-2}$ s$^{-1}$) against the power-law normalization (units of ph. keV$^{-1}$ cm$^{-2}$ s$^{-1}$ at 1 keV). The 1989 data are from Ginga, May 1995 and 1996 are from ASCA. October 1996 and November 1997 are the BeppoSAX data presented in this paper.}
\end{figure}

\begin{figure}[h]
\psfig{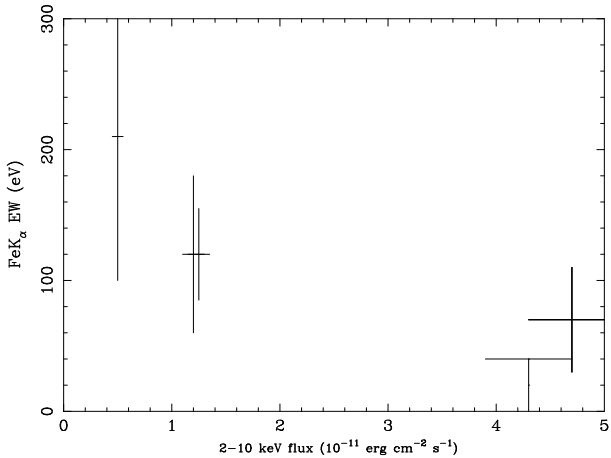}
\caption{The FeK$\alpha$ line EW against the 2-10 keV flux during the last 9 years' observations (data are, from left to right, from BeppoSAX 1997, BeppoSAX 1996, ASCA 1996, Ginga 1989, ASCA 1995).}
\end{figure}

To further investigate the issue of the FeK$_{\alpha}$ line variability, we
reanalyzed the old Ginga and ASCA observations of NGC 7172 
in order to estimate the order of magnitude of the changes involved. 

All the data were fitted with a simple absorbed power 
law plus FeK$\alpha$ line so as to have uniformity in the treatment of the 
continuum modeling (note that the reflection component is not required by the
ASCA data). In figure 10 the line flux is plotted against the 1 keV continuum 
normalization, while in figure 11 the line equivalent width is shown as a 
function of the 2-10 keV flux.

If the line originates in the accretion disk, then the change by a factor of
$\sim$10 in the continuum level (figure 10), would produce a 
change of an equal amount in the line intensity, thus providing a constant EW
in time.

Instead, we observe a decrease in the line intensity of no more than a factor 
of $\sim$7 (this value was obtained in figure 10 considering the 90\% 
confidence contours). Moreover, figure 11 shows an anti-correlation between 
the line EW and the 2-10 keV flux. We can therefore conclude that the data are
not compatible with a pure disk origin of the line.

If instead the line is produced in the absorber itself, 
assuming a spherical distribution, we would expect an EW $\sim$70 eV for the 
observed N$_{H}$ (Makishima 1986). Otherwise, in the framework of the unified 
models for AGN (Antonucci \& Miller 1985, Antonucci 1993) a molecular torus seen almost edge-on, with the observed 
N$_{H}$, would produce an FeK$\alpha$ line with 
an EW of 30-40 eV (Ghisellini, Haardt \& Matt 1994). 
This value is in perfect agreement with 
what was measured by Ginga (Nandra \& Pounds 1994) and during the 1995 ASCA 
observation (Ryde et al. 1996, 1997) when the source was bright but somewhat
lower than observed by BeppoSAX. The agreement could, however, be recovered 
and 
the anti-correlation in figure 11 could  be explained if there is a  
time lag between the continuum and the line production region: 
as the flux 
level of the nucleus decreases, the line intensity follows these changes with 
a delay directly proportional to the distance between the continuum 
production region and the reprocessor. The overall  effect is that the line 
EW increases as the source weakens. Assuming this 
simplified model we have the opportunity to calculate a lower limit for the 
distance between the emitter and the reprocessor which is of the order of 
$\sim$8 light years for NGC 7172  (d$\leq$c$\times$$\Delta${\it t} where 
d is the linear dimension, c is the speed of light and $\Delta${\it t} is the 
observed time scale of the variations).

\section{Discussion and conclusions}

We analyzed
two BeppoSAX observations, performed in October 1996 and November 1997, of the
sky region containing NGC 7172. The source is detected embedded in diffuse 
soft-X emission from the compact group HCG90, of which the Seyfert 2 is a 
member.

NGC 7172 showed flux variability in the 2-10 keV range. During 
observation B the source flux varied by a factor of $\sim$30\% thus indicating 
the presence of a type 1 nucleus in its center, while the luminosity decreased 
by a factor of $\sim$2 between the two BeppoSAX pointings.
Indeed, we detected the
lowest flux level ever recorded for this source in the 2-10 keV band. During
the November 1997 pointing, the source was $\sim$10 times weaker than
when observed a decade before.

A good fit to the data of both observations in the whole 2-90 keV range is 
yielded by a simple power-law plus Gaussian line model, photoelectrically 
absorbed by a substantial column density of cold matter. The flat observed 
spectrum ($\Gamma$=1.6-1.7) could be due to the presence of a Compton 
reflection component. The presence of this component is indeed required by 
the fit at a more than 90$\%$ confidence level, if we want the intrinsic 
power-law index to be equal to the canonical value, 1.9, observed in Seyfert 
galaxies. In this case, the amount 
of reflection is constrained only in the range $\sim$1-3 (weighted mean of the BeppoSAX measurements). However, the systematics introduced by the uncertinities in the relative normalization between MECS and PDS are likely to substantially affect this (Fiore Guainazzi \& Grandi 1999).

The absorbing column density is  N$_{H}$$\sim$10$^{23}$cm$^{-2}$ 
(independent of the model adopted for the description of the continuum), as 
previously observed. This value
seems to have remained stable over a $\sim$20 years period 
(Malizia et al. 1997, Turner \& Pounds 1989, Smith \& Done 1996, 
Guainazzi et al. 1998) although the source has dimmed by a factor of $\sim$10.

A complex absorption model was also tested  in order to check if this is 
the cause of the flatness of the spectrum. The fit obtained is 
statistically acceptable but we discarded this model since it is not required 
by the data.

X-ray observations of Seyfert galaxies support the idea that the primary X-ray 
continuum is reflected from cold material surrounding the central nucleus 
(Lightman \& White 1988,  Ghisellini, Haardt \& Matt 1994); this  
reprocessing produces an FeK$\alpha$ line and a reflection component. 
Whether this occurs in the accretion disk or
in other structures of the nuclear region (for example the molecular 
torus of the unified models for AGN) is still debatable
for Seyfert 2 galaxies. Given the large amplitude of NGC 7172 continuum 
variations seen in the last 10 years, it is possible to tackle this issue in 
this particular case,  at least for the FeK$\alpha$ line. We therefore 
searched the data for evidence of possible changes in the line parameters so 
as to understand its origin and location. 

The pure disk origin of the line is ruled out from the present data for two 
main reasons: 1) the line intensity does not follow the continuum instantly;
2) the equivalent widths obtained by Ginga ($\leq$40 eV) and ASCA 1995 
observations ($\sim$60 eV) are lower than expected (EW $\sim$100-200 eV) 
from a disk in a face-on configuration. To reconcile these low EW values with 
a pure disk origin, one should assume that the disk is highly inclined. In this
case we expect that the absorber would contribute (with the  
N$_{H}$$\sim$10$^{23}$cm$^{-2}$ seen in NGC 7172) by a similar 
amount to the line EW (Makishima 1986, Ghisellini, Haardt \& Matt 1994), 
overproducing what is observed.

We therefore conclude that the FeK$\alpha$ line is most likely 
produced far from the continuum source and that it probably emerges from the 
absorber itself.
Here the fluorescent line would be produced via transmission in an
optically thin medium. In this case no reflection component would be 
associated with the FeK$\alpha$, in agreement with the present measurements.
The time lag between the continuum changes and the line reaction allows us to 
infer the size of the absorber, which is of the order of $\sim$8 light years. 

We speculate that such an absorber can be associated with a parsec-scale 
molecular torus, as usually assumed in standard unified models for 
AGN (Antonucci \& Miller 1985, Antonucci 1993).
This interpretation would also be in agreement with the EW measured by Ginga 
and ASCA (Smith \& Done 1996, Guainazzi et al. 1998). 

\vspace{1.0cm}

After the present work was completed, we became aware of a paper discussing 
the same BeppoSAX sets of data by Akylas et al. (2000). The spectral fits 
obtained by Akylas et al. (2000) are 
consistent with the present results, although the authors prefer 
to interpret the data in other terms.

\begin{acknowledgements}

The authors want to thank the anonymous referee for his/her helpful comments 
and patience. This work has been partially supported by the Italian Space Agency 
(ASI).

\end{acknowledgements}

\end{document}